\documentclass[journal=nalefd,manuscript=letter]{achemso}

\usepackage[utf8]{inputenc}
\usepackage{polski}
\usepackage{graphicx}
\usepackage{dcolumn}
\usepackage{bm}
\usepackage{hyperref}
\usepackage{color}
\usepackage[justification=justified,
   format=plain]{caption}
\usepackage{subcaption}
\usepackage{amsmath}
\usepackage{amssymb}
\usepackage{amsthm}
\usepackage{csquotes}
\usepackage{amsfonts}
\usepackage{physics}
\usepackage[english]{babel} 
\usepackage[version=3]{mhchem}

\title{Magneto-Optics of Anisotropic Exciton Polaritons in Two-Dimensional Perovskites}
\author{Jonas K. König}
\affiliation[Marburg]{Fachbereich Physik, Philipps-Universit\"at, Marburg, 35032, Germany}
\email{jonas.koenig@physik.uni-marburg.de}
\author{Jamie M. Fitzgerald}
\affiliation[Marburg]{Fachbereich Physik, Philipps-Universit\"at, Marburg, 35032, Germany}

\author{Ermin Malic}
\affiliation[Marburg]{Fachbereich Physik, Philipps-Universit\"at, Marburg, 35032, Germany}
\keywords{exciton polaritons; dark excitons;  anisotropic polaritons; magneto polaritons; 2D perovskites}

\begin{document}

\maketitle
\begin{abstract} 
Layered 2D organic–inorganic perovskite semiconductors support strongly confined excitons that offer significant potential for ultrathin polaritonic devices due to their tunability and huge oscillator strength. The application of a magnetic field has proven to be an invaluable tool for investigating the exciton fine structure observed in these materials. Yet, the combination of an in-plane magnetic field and the strong coupling regime has remained largely unexplored. In this work, we combine microscopic theory with a rigorous solution of Maxwell's equations to model the magneto-optics of exciton polaritons in 2D perovskites. We predict that the brightened dark exciton state can enter the strong coupling regime. Furthermore, the magnetic-field-induced mixing of polarization selection rules and the breaking of in-plane symmetry lead to highly anisotropic polariton branches. This study contributes to a better understanding of the exciton fine structure in 2D perovskites and demonstrates the cavity control of highly anisotropic and polarization-sensitive exciton polaritons. 
\end{abstract}

\maketitle

\section*{Introduction}

Layered 2D hybrid organic-inorganic metal halide perovskites have attracted a great deal of interest in recent years due to their improved environmental stability and superior tunability compared to their conventional 3D counterparts \cite{pedesseau2016advances,chen20182d}. They offer potential applications in ultrathin light emitters \cite{dou2015atomically,hu2016molecularly}, photovoltaics \cite{tsai2016high,chen2017tailoring}, photodetectors \cite{zhou2016photodetectors}, and chiral optoelectronics \cite{liu2023bright}. Composed of an inorganic perovskite layer sandwiched between two layers of organic spacers acting as potential barriers, they form intrinsic 2D quantum well heterostructures \cite{stoumpos2016ruddlesden}. These naturally stack on top of one another to form a single crystal slab, significantly enhancing the light-matter interaction through collective effects without modifying electronic properties \cite{fieramosca2018tunable}, in contrast to transition metal dichalcogenides \cite{liu2015strong}. Their remarkable optical properties are governed by tightly bound excitons confined to the plane of the inorganic layer \cite{blancon2018scaling}, and exhibit a rich exciton fine structure that comprises bright triplet and dark singlet states \cite{tanaka2005electronic,baranowski2020excitons}. Application of a twist angle \cite{zhang2024moire} and modification of the organic spacer \cite{dyksik2024steric} have recently been shown to provide a tunable exciton oscillator strength and spacing, respectively. Emission studies of both nanocrystal and 2D lead-iodide perovskites have revealed deviations of the exciton distribution from Boltzmann statistics, resulting in surprisingly intense emission from higher-energy bright states even at cryogenic temperatures \cite{xu2018long,tamarat2019ground,dyksik2021brightening}. This is a direct consequence of an exciton relaxation bottleneck caused by a mismatch between the dark-bright exciton splitting and the energy of the involved optical phonons \cite{thompson2024phonon}. There has been controversy in the literature regarding the energetic ordering of the dark and lowest bright states \cite{becker2018bright,lou2024robust}. The application of magneto-optical spectroscopy has played a key role in this debate, as it enables the direct observation of the dark state through the mixing of the singlet with a neighboring bright triplet state \cite{tamarat2019ground,tamarat2020dark,dyksik2021brightening,posmyk2022quantification,thompson2024phonon}.  

\begin{figure}[t]
    \centering
    \includegraphics[width=0.5\textwidth]{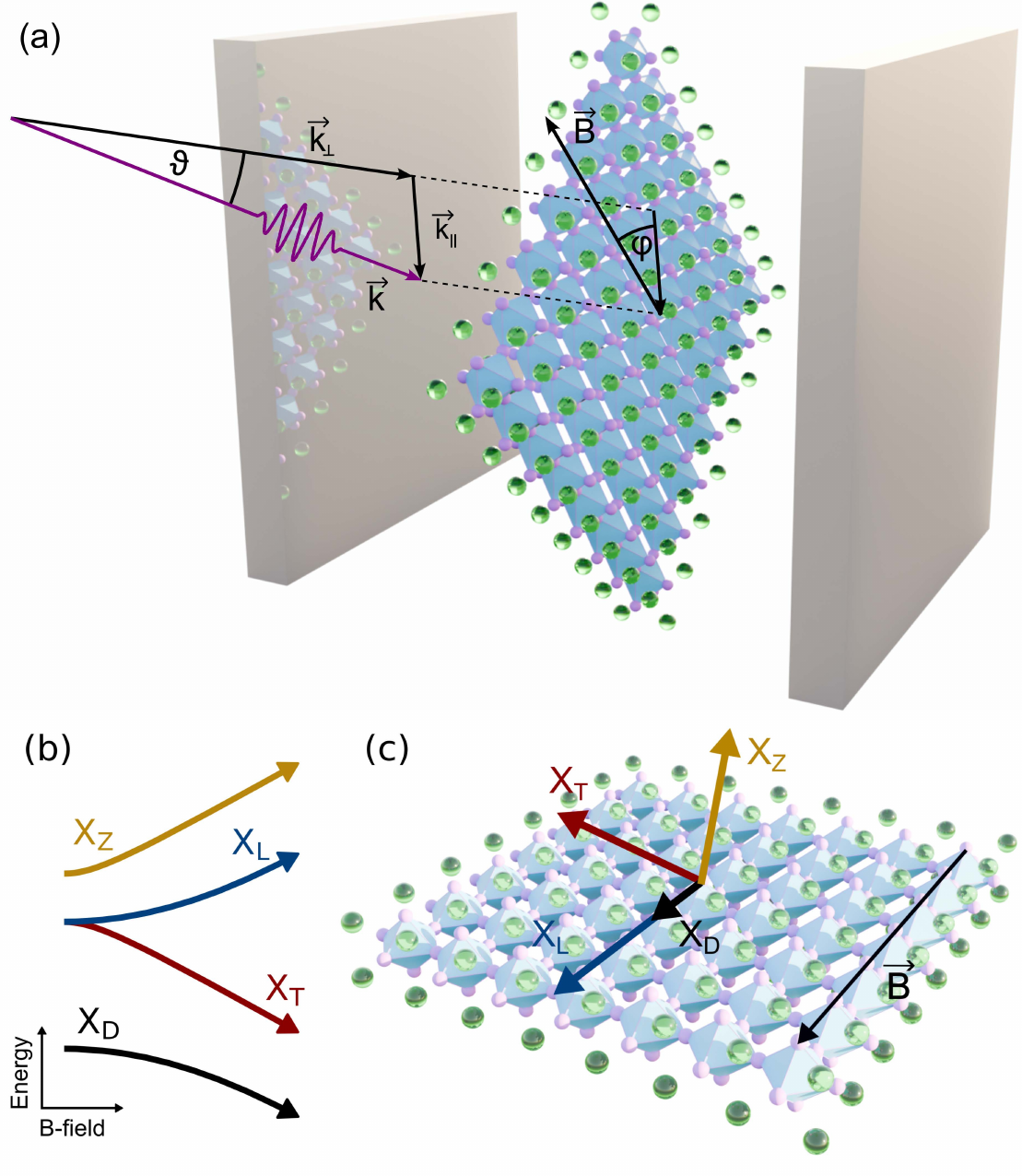}
    \caption{(a) Schematic figure of a 2D PEA$_2$PbI$_4$ perovskite slab integrated within a Fabry-P\'erot cavity. The photon (purple arrow) is incident at an angle $\vartheta$ and momentum $\textbf{k}$, where $\textbf{k}_\perp(\textbf{k}_\parallel)$ denote the perpendicular (parallel) component with respect to the perovskite layer. There is an in-plane magnetic field $\textbf{B}$ with an azimuth angle $\varphi$ relative to $\textbf{k}_\parallel$. (b) Magnetic field dependence of the four exciton fine structure states: the dark ($X_D$), the two bright ($X_T$, $X_L$), and the gray exciton ($X_Z$). (c) Direction of the corresponding transition dipole moment for the four states in the presence of a magnetic field. While $X_D$ and $X_L$ are longitudinally polarized, $X_T$ and $X_Z$ are transversally polarized with respect to the magnetic field. The arrow length denotes the respective magnitude of the oscillator strength.}
   \label{fig:schematic}
\end{figure}

The large oscillator strength of their bright excitons, resulting from quantum and dielectric confinement, makes 2D metal-halide perovskites exceptional candidates for room-temperature exciton polaritonics \cite{su2021perovskite}. The strong coupling regime, with Rabi splittings in the range of hundreds of meV, has been achieved for perovskites integrated as active layers with planar microcavities \cite{lanty2008strong,wang2018room,fieramosca2019two,laitz2023uncovering}, plasmonic structures \cite{symonds2007emission,niu2015image}, as well as cavity-free self-hybridized slabs \cite{fieramosca2019two,anantharaman2021self,anantharaman2024dynamics} and photonic crystals/metasurfaces \cite{masharin2023room,wu2024exciton,dang2024long}. Interesting cavity physics has already been explored, including polariton bottlenecks \cite{laitz2023uncovering}, topological polaritons \cite{su2021optical,jin2024observation}, the optical spin Hall effect \cite{shi2024coherent}, and polariton lasing \cite{masharin2023room} and condensation \cite{su2020observation,wu2024exciton}. In addition, combining polaritonics with the application of a magnetic field has been used to tune the Berry curvature of exciton polariton bands \cite{polimeno2021tuning}, and, in general, provides interesting opportunities for anisotropic polaritonics \cite{wang2024planar} via the breaking of in-plane symmetry. The brightening of the dark fine structure exciton \cite{dyksik2021brightening} makes it relevant for photonics, offering the potential for cavity control of spin \cite{lu2020highly}, polarization \cite{posmyk2022quantification,wang2022electrically}, and directional transport \cite{voronin2024fundamentals,dang2024long}: crucial ingredients for many quantum optoelectronic applications \cite{tamarat2020dark}. The photonic hybridization of different exciton states, both with and without an applied magnetic field, presents an exciting strategy to control, characterize, and visualize the still-debated exciton fine structure in 2D perovskites.

Based on a microscopic, material-specific, and predictive many-particle theory, we investigate the exciton fine structure of exemplary layered (PEA)$_2$PbI$_4$ perovskites \cite{hong1992dielectric}, consisting of a \emph{single} lead-iodide perovskite layer sandwiched between two layers of phenylethylammonium as organic spacers. We combine the Wannier equation, which provides microscopic access to exciton characteristics, with a full-wave solution of Maxwell's equations to describe exciton polaritons in 2D perovskites integrated within a Fabry-P\'erot microcavity. Employing an in-plane magnetic field gives rise to rich optical selection rules, brightening the dark state \cite{dyksik2021brightening} and even potentially enabling it to enter the strong coupling regime. We study the effects of the magnetic field in terms of Rabi splitting and absorption of the polariton landscape. In particular, we observe interesting superimposed anisotropic polariton branches due to the magnetic field breaking the in-plane symmetry. Our work provides a first prediction of the magneto-absorption of exciton polaritons in 2D perovskites formed from the brightened dark state, highlighting their experimental signatures and tunability using the magnetic field strength and cavity length. This has relevance for potential ultrathin and tunable polarization-sensitive photonic devices, as well as for the characterization and visualization of the still-unresolved exciton fine structure in 2D perovskites.

\section*{Results and discussions}

\subsection*{Theoretical Approach}
To obtain the exciton energy landscape, we first solve the Wannier equation \cite{thompson2024phonon}, which provides microscopic access to excitonic wavefunctions and binding energies. The short- and long-range exchange interaction between the electrons and holes is then converted to the exciton picture, and the resulting Hamiltonian is diagonalized (see SI for further details). This results in an exciton fine structure energy landscape, including rich optical selection rules \cite{thompson2024phonon}. With this approach, we are able to accurately describe previously observed features of the exciton fine structure of a 2D (PEA)$_2$ PbI$_4$ perovskite layer \cite{dyksik2021brightening,posmyk2022quantification,posmyk2024exciton}. To model the optics of the perovskite layer, both in vacuum and within a Fabry-P\'erot microcavity, we solve Maxwell's equations using the scattering matrix (S-matrix) method, which is suitable for layered media that are spatially homogeneous in the plane \cite{rumpf2011improved} (see SI). Excitons and their selection rules are included via a dispersive and anisotropic dielectric tensor
\begin{equation}
    \varepsilon(\omega)=\begin{pmatrix}
        \varepsilon_B+\chi_{x}(\omega) & 0 & 0\\
        0 & \varepsilon_B+\chi_{y}(\omega) & 0\\
        0 & 0 & \varepsilon_B
    \end{pmatrix}\;\text{,}
\end{equation}
with $\varepsilon_B$ being the dielectric background of the perovskite, and $\chi_{x(y)}$ the frequency-dependent response function of the material in the in-plane x(y)-direction. The latter is given by \cite{kira2006many}
    \begin{align}
        \chi_{x(y)}&
        \propto \sum_\mu\frac{\hbar\gamma_{\mu}^{x(y)}(\vartheta=0^\circ)}{(E_\mu-\hbar\omega)-i\hbar\Gamma_\mu}\;\text{,}
    \end{align}
where $E_\mu$ is the energy of the $\mu$th exciton and $\hbar\Gamma_\mu$ the corresponding exciton scattering rate. $\hbar\gamma_\mu^{x(y)}$ is the corresponding exciton radiative decay in the $x(y)$-direction, which determines the oscillator strength (Fig. S1 in the SI). Changing the momentum components of the incoming photon allows us to vary both the angle of incidence, $\vartheta$, and the azimuth angle, $\varphi$, with respect to the magnetic field, as shown in Fig. \ref{fig:schematic}(a). For simplicity, we focus only on TE-polarized light, i.e., where the polarization of the light is purely in the $xy$-plane, and the excitonic contribution from the out-of-plane $z$-component can be ignored.

We consider a microcavity consisting of a pair of Bragg mirrors, with a single perovskite quantum well in the center (Fig.\ref{fig:schematic}). Using the S-matrix method, with microscopic input from the Wannier equation, we calculate the linear optical spectra of the combined system. While this provides exact solutions to Maxwell's equations, it does not give access to the Hopfield coefficients, which are necessary for a detailed understanding of the constituent nature and decay channels of each polariton \cite{fitzgerald2022twist,ferreira2022signatures}. We therefore extract the exciton polariton energies from the calculated reflection, and the resulting dispersion manifolds are then fitted to a two-exciton-one-photon Hopfield model \cite{hopfield1958theory,fitzgerald2022twist}. This provides microscopic access to the cavity photon-exciton coupling strength, $g_\mu$, and the Hopfield coefficients, $U_\mu^n$. The Hopfield matrix is given by 
\begin{align}
    H&=\begin{pmatrix}
        E^{(C)} & g_1 & g_2\\
        g_1 & E^{(X)}_1 & 0\\
        g_2 & 0 & E^{(X)}_2
    \end{pmatrix}\;\text{,}\label{eq:hopfield}
\end{align}
where $E^{(X)}_\mu$ is the energy of the $\mu$th exciton and $E^{(C)}$ denotes the energy of the bare cavity photon, which is extracted from an S-matrix simulation ignoring excitonic effects.

To further analyze the results, we use coupled mode theory, which is appropriate for describing the coupling of low-loss material resonances and high-Q photonic modes to each other and to external ports \cite{fan2003temporal} (see the SI). The absorption of a single resonator (valid for excitons or polaritons) is given by the Elliott formula \cite{kira2006many,fitzgerald2022twist}
\begin{equation}
    A(\omega)=\frac{2\hbar\Gamma\hbar\gamma}{(\hbar\omega_R-\hbar\omega)^2+(\hbar\Gamma+\hbar\gamma)^2}\;\textrm{,} \label{eq:elliot}
\end{equation}
with $\omega_R$ being the frequency of the respective resonator, while $\hbar\gamma$ and $\hbar\Gamma$ are the photonic and material-based decay channels, respectively. For a symmetric system excited from one port, the absorption has an upper bound \cite{piper2014total} of $0.5$, which is reached at the critical coupling condition \cite{ferreira2022signatures,konig2023interlayer} of $\gamma=\Gamma$, as can be seen from Eq. (\ref{eq:elliot}).

\subsection*{Exciton fine structure and magneto-absorption}
We use the approach presented above to investigate the exciton fine structure and absorption in 2D perovskites under an applied in-plane magnetic field, both with and without integration into a Fabry-P\'erot cavity. Solving the Wannier equation, we obtain a binding energy of approximately 220\,meV for the four-fold degenerate exciton, originating from the possible spin configurations of electrons and holes \cite{baranowski2020excitons}: in excellent agreement with prior theoretical \cite{feldstein2020microscopic,thompson2024phonon} and experimental \cite{hong1992dielectric,urban2020revealing} studies. Including the exchange interaction, we find that the four degenerate spin states split in energy. The energetically lowest singlet state is optically dark \cite{tamarat2019ground,dyksik2021brightening} and is therefore labeled the dark state, $X_D$. Furthermore, there are two degenerate circularly polarized bright states, $X_{T/L}$, and an out-of-plane polarized gray state, $X_Z$ \cite{fieramosca2018tunable,posmyk2024exciton}. If an in-plane magnetic field is applied (Voigt configuration), these states further mix, modifying their energy and optical selection rules \cite{dyksik2021brightening}. Their energy shift as a function of the magnetic field strength is sketched in Fig. \ref{fig:schematic}(b). Level repulsion causes the dark state to shift down in energy, the gray exciton to shift up, and the two degenerate bright states to split apart. The previously circularly polarized states become linearly polarized \cite{posmyk2022quantification}, where $X_T$ is orthogonal (transversal) to the magnetic field but no longer fully in-plane, while $X_L$ is parallel (longitudinal). Furthermore, the dark state brightens and becomes polarized along the magnetic field direction due to field-induced mixing with the $X_L$ state as a consequence of the spin selection rules \cite{feierabend2020brightening}. Lastly, the gray state is also transversally polarized, but it is no longer strictly orientated out-of-plane due to mixing with the $X_T$ state. The orientation of the transition dipole moments with respect to the $B$-field is illustrated in Fig. \ref{fig:schematic}(c).

\begin{figure}[t]
    \centering
    \includegraphics[width=1\textwidth]{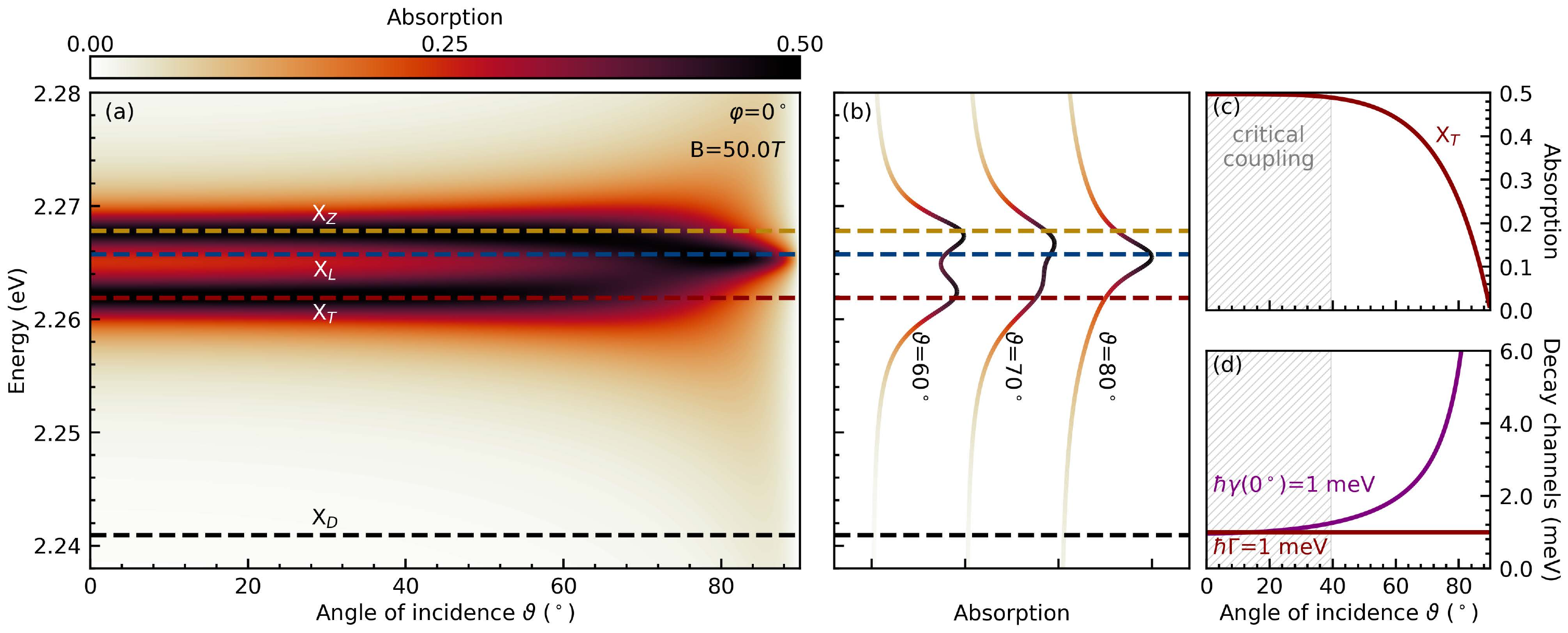}
    \caption{(a) Absorption of a 2D PEA$_2$PbI$_4$ perovskite layer as a function of photon energy and angle of incidence for $B=50\,\mathrm{T}$ and $\varphi=0^\circ$. The horizontal dashed lines indicate the energies of the four exciton fine structure states. As we consider TE-polarized light, i.e., a polarization perpendicular to the magnetic field at $\varphi=0^\circ$, only $X_T$ and $X_Z$ excitons can couple to light at this orientation, as shown in Fig. \ref{fig:schematic}(c). (b) Line cuts of the absorption spectrum at specific higher angles of incidence. (c) Absorption along the energy of $X_T$. (d) Radiative ($\hbar\gamma$) and material-based ($\hbar\Gamma$) decay channels for $X_T$. The shaded area indicates where the two decay channels have a relative difference of less than $5\%$, marking the critical coupling region with maximal absorption.}
   \label{fig:bare_perov}
\end{figure}

According to these selection rules, when the bare perovskite slab is excited with TE-polarized light in the absence of a magnetic field, the response is independent of the azimuth angle $\varphi$ and only a single absorption peak at the energy of the degenerate bright states is observed. In contrast, in the presence of a high magnetic field, the optical response depends crucially on $\varphi$. For $\varphi=0^\circ$ (i.e., when the electric field of the light is fully in-plane and orthogonal to the applied magnetic field), only the transversal states, $X_T$ and $X_Z$, can couple to the light via their in-plane component \cite{dyksik2021brightening}. Considering a magnetic field strength of 50\,T, this leads to two absorption peaks at low angles of incidence, as illustrated in Fig. \ref{fig:bare_perov}(a). For higher angles of $\vartheta$, the two peaks merge and there is a strong absorption approximately at the $X_L$ energy. A closer analysis of the two peaks reveals that they broaden and overlap, creating the impression of a single peak between them that coincides with $X_L$, as illustrated in Fig. \ref{fig:bare_perov}(b). For TE-polarized light, the radiative decay $\hbar\gamma_\mu$ of a given exciton state $\mu$ can be shown to scale with \cite{kavokin1995excitionic} $1/\cos(\vartheta)$ and thus diverges as $\vartheta$ limits towards 90$^\circ$  \cite{creatore2008strong}. Consequently, the full width at half maximum of the peak, given by $2(\hbar\Gamma_\mu+\hbar\gamma_\mu)$, also increases rapidly as $\vartheta$ approaches grazing angles. Surprisingly, as the radiative decay of the exciton diverges, the absorption decreases to zero, as shown in Fig.~\ref{fig:bare_perov}(c). This occurs because the difference between radiative and material-based decay rates increases with $\vartheta$ (Fig.~\ref{fig:bare_perov}(d)), moving the system away from the critical coupling regime and therefore decreasing the absorption.

\subsection*{Polariton magneto-absorption in an optical cavity}
\begin{figure}[t!]
    \centering
    \includegraphics[width=0.5\textwidth]{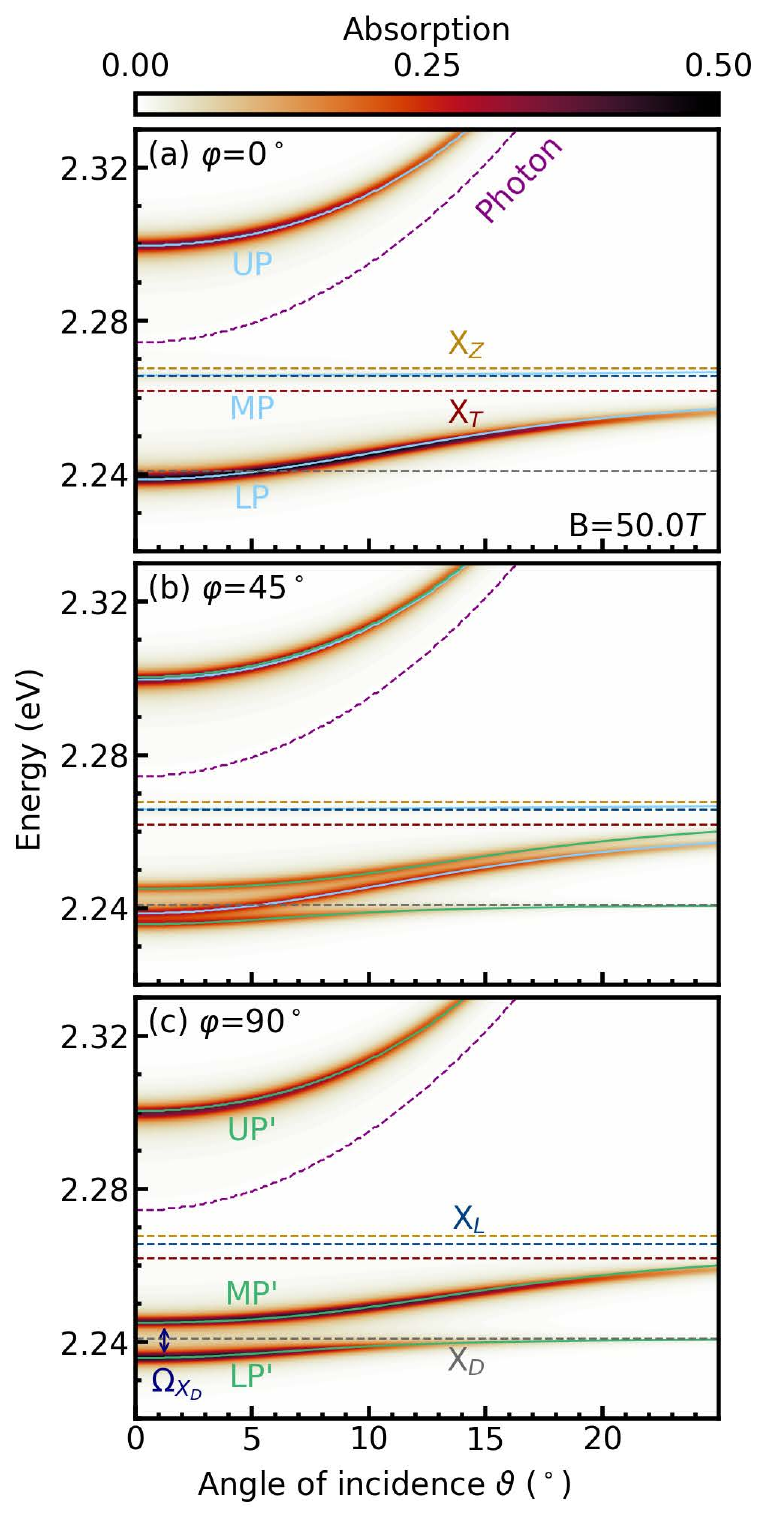}
    \caption{(a)-(c) Absorption of a 2D (PEA)$_2$PbI$_4$ perovskite layer integrated within a Fabry-P\'erot microcavity as a function of the photon energy and angle of incidence for $B=50\,\mathrm{T}$ and different azimuth angles $\varphi$. The horizontal dashed lines indicate the exciton energies. Depending on $\varphi$, different polariton branches (denoted by LP, MP, and UP for $\varphi=0^\circ$ and LP', MP', and UP' for $\varphi=90^\circ$) appear as a result of the selection rules of the respective excitonic states. The small vertical arrow in part (c) denotes the Rabi splitting of the dark state, $\Omega_{X_D}$, which brightens in the presence of a magnetic field.}
   \label{fig:perov_FP}
\end{figure}

The integration of a perovskite slab into a microcavity, combined with the brightening of the dark state in the presence of an in-plane magnetic field, presents a potential strategy to investigate the still-debated energetic ordering of the exciton fine structure states in these materials. Furthermore, the anisotropy induced by the magnetic field is expected to have a dramatic impact on polariton optics. 
In particular, by varying the azimuth angle $\varphi$, the projection of the electric field onto the transition dipole moment of the various fine structure states can be altered relative to the applied in-plane magnetic field orientation, leading to a rich anisotropic polariton dispersion that can be selectively probed with the incident beam angle. The polariton landscape in the absence of the magnetic field is discussed in the SI. Here, we focus on the case of magneto-optics. Starting with $\varphi=0^\circ$, where the electric field is orientated perpendicular to the applied magnetic field, only the two excitons, $X_T$ and $X_Z$, can couple to the incident laser, similar to the bare perovskite case. This results in three polariton branches, a lower (LP), middle (MP) and upper polariton branch (UP), cf. Fig.\ref{fig:perov_FP}(a). As the middle branch is sandwiched between the almost energetically degenerate states $X_T$ and $X_Z$, it is nearly flat and therefore barely visible in absorption \cite{fitzgerald2022twist}. As a result, only one large apparent Rabi splitting is observed, instead of the two splittings typically observed for a two-exciton, one-photon system. However, the absorption of the middle branch increases with stronger magnetic fields as the separation between $X_T$ and $X_Z$ grows.

For $\varphi=90^\circ$, where the electric field is orientated parallel to the applied magnetic field, only $X_L$ and $X_D$ can couple to the light. This leads again to three polariton branches (LP', MP', UP'), cf. Fig.~\ref{fig:perov_FP}(c). As $X_L$ and $X_D$ are well separated in energy, the middle polariton is no longer flat, and we observe two Rabi splittings: one around $X_L$ and a smaller splitting around $X_D$. The latter is labeled as $\Omega_{X_D}$ and defined as the difference between LP' and MP' at $\vartheta=0^\circ$. The upper polariton branch (UP) is approximately independent of $\varphi$ and hence very similar for the $0^\circ$ and $90^\circ$ cases. This is because the combined oscillator strength of $X_Z$ and $X_T$ is approximately equal to that of $X_L$, due to the conservation of oscillator strength in the presence of Zeeman splitting \cite{thompson2024phonon}.

Applying a fit using the Hopfield model (Eq. (\ref{eq:hopfield})) gives a Rabi splitting of $\Omega_{X_D}=9\,\mathrm{meV}$ for the dark state at $B=50$\,T, while for $X_L$ the splitting is 54\,meV. In comparison, the Rabi splitting of the $X_T$ and $X_Z$ states in the $\varphi=0^\circ$ case is 60\,meV. The detuning was chosen such that the LP in Fig.~\ref{fig:perov_FP}(a) coincides approximately with the dark state, leading to a Rabi splitting around the dark state, $\Omega_{X_D}$, to be at $\vartheta=0^\circ$, as indicated by the small vertical arrow in Fig.~\ref{fig:perov_FP}(c). Smaller cavity lengths would increase the energy of the cavity photon mode, thereby detuning the cavity from the $X_D$ energy and leading to a smaller Rabi splitting. In contrast, a longer cavity length would decrease the cavity photon energy, shifting the Rabi splitting to larger angles of incidence. These results constitute a first prediction that, at certain incident laser orientations, the brightened low-energy dark state can enter the strong coupling regime and be observed by magneto-optical measurements. This paves the way towards cavity control of the exciton fine structure in 2D perovskites.

Interestingly, at an intermediate angle of $\varphi=45^\circ$, we do not observe a smaller Rabi splitting around the brightened dark state (Fig. \ref{fig:perov_FP}(b)), as might be naively expected from considering only the projection of the oscillator strength. Instead, we observe a superposition of the two previously discussed limiting cases of $\varphi=0^\circ$ and $\varphi=90^\circ$, with unchanged Rabi splittings but altered absorption. The upper branch still has the same absorption due to the near-degeneracy of UP and UP', while the other branches have their absorption approximately halved. This behavior of superimposed Rabi splittings is related to a magnetic-field-induced anisotropy of the dielectric tensor, which arises from the two inequivalent dipoles pointing in different directions. Similar results have previously been observed in orientated molecular aggregates \cite{balagurov2004organic} and 2D naturally anisotropic TMD (ReS$_2$) layers \cite{chakrabarty2021interfacial}. In the Hopfield model, this means that there are two degenerate cavity modes polarized orthogonal to each other, which oscillate perpendicular (parallel) to the magnetic field and couple only to $X_T$ and $X_Z$ (only to $X_D$ and $X_L$). Therefore, this six-branch system decouples into two three-branch subsystems, orientated perpendicular and parallel to the magnetic field, respectively, and exhibiting the same cavity-exciton couplings as before. The observed magnetic field and cavity control of the polariton landscape, particularly the modification of the Rabi splitting relative to typical phonon energies, presents opportunities to manipulate optics \cite{feldstein2020microscopic,ferreira2024revealing} and relaxation in these materials \cite{thompson2024phonon,fitzgerald2024circumventing}, with relevance, e.g., to polariton lasing \cite{butte2002transition}. The anisotropic polariton bands will lead to a direction-dependent relaxation, with the middle polariton branch anticipated to play an important role in dictating relaxation from higher-energy states to the lower polariton branch \cite{hu2024energy}.

\subsection*{Tunability of Rabi splitting and absorption intensity}
\begin{figure}[t]
    \centering
    \includegraphics[width=0.6\textwidth]{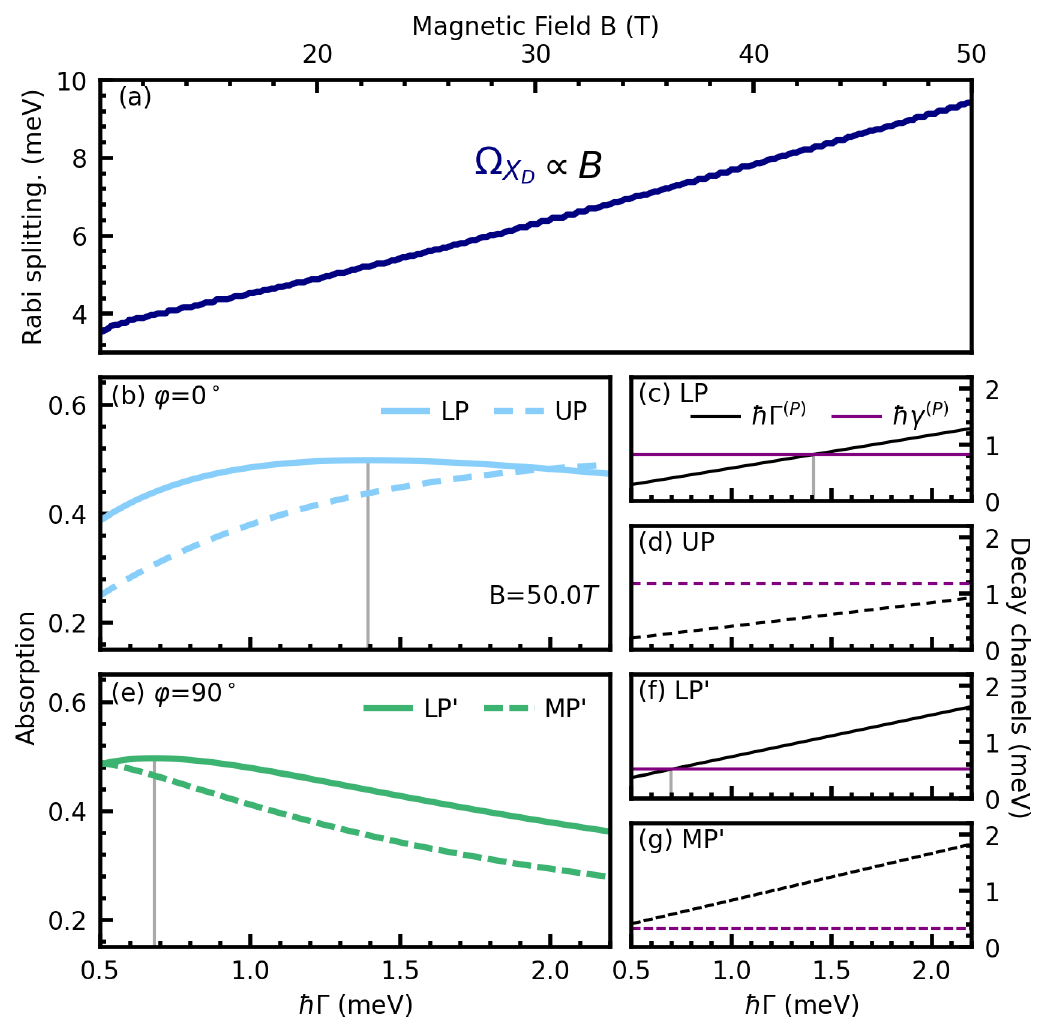}
    \caption{(a) Rabi splitting, $\Omega_{X_D}$, stemming from the avoided crossing between the dark state and the photon energy, as a function of the magnetic field strength $B$ (see the arrow in Fig.~\ref{fig:perov_FP}(c)). (b) Resonant absorption of the lower (LP) and upper (UP) polariton branches from \ref{fig:perov_FP}(a) ($\varphi=0^\circ$) at $\vartheta=0^\circ$ as a function of the excitonic linewidth $\hbar\Gamma$. The vertical gray line indicates the point where the absorption reaches a peak value of 0.5. (c)-(d) Photonic, $\hbar\gamma^{(P)}$, and excitonic, $\hbar\Gamma^{(P)}$, decay channels of the polariton branches LP and UP, respectively. The vertical gray line shows where the two are equal, i.e., the critical coupling condition is met. It coincides exactly with the vertical gray line in (b). (e) The same plot as in (b), but for $\varphi=90^\circ$, illustrating the lower (LP') and the middle (MP') polariton branches from Fig.~\ref{fig:perov_FP}(c). (f)-(g) Photonic and excitonic decay channels of the corresponding polariton branches LP' and MP', respectively.}
   \label{fig:perov_tuning}
\end{figure}

Now, we explore the tunability of the system with respect to the applied in-plane magnetic field. With increasing field strength, oscillator strength is transferred from $X_L$ to $X_D$ (see SI). Therefore, the Rabi splitting $\Omega_{X_D}$ around the dark state is also expected to grow. We find that $\Omega_{X_D}$ scales linearly with the magnetic field, cf. Fig.~\ref{fig:perov_tuning}(a). The oscillator strength of the dark state scales quadratically with the magnetic field, provided that the Zeeman splitting $\mu_B B$ is small compared to the energy difference between the $X_L$ and $X_D$ states \cite{feierabend2020brightening}. Furthermore, the coupling between exciton and cavity photon modes is proportional to the square root of the oscillator strength \cite{fitzgerald2022twist}, explaining the observed linear behavior. The absorption as a function of the magnetic field strength is provided in the SI.

Another way to tune the system is to change the exciton linewidth $\hbar\Gamma$ via, e.g., varying the temperature \cite{feldstein2020microscopic}. The material-based loss of a polariton branch $n$ is given by the sum of the linewidths of all its excitonic components, weighted by the respective excitonic Hopfield coefficients $\hbar\Gamma^{(P)}_n=\sum_\mu\hbar\Gamma\left(U_\mu^n\right)^2$\cite{fitzgerald2022twist,ferreira2022signatures}. Similarly, the radiative decay is given by the cavity mode linewidth, $\hbar\kappa$, scaled by the photonic Hopfield coefficient $\hbar\gamma^{(P)}_n=\hbar\kappa\left(U_0^n\right)^2$. The different branches in Fig.~\ref{fig:perov_FP}(a) reach a large absorption at $\vartheta=0^\circ$ for different $\hbar\Gamma$, as illustrated in Fig. \ref{fig:perov_tuning}(b). For a particular polariton branch, we find that high absorption coincides with regions where the respective excitonic and photonic loss rates are similar (Figs. \ref{fig:perov_tuning}(c)-(d)), i.e., fulfilling the critical coupling condition that is contained in Eq. (\ref{eq:elliot}) \cite{fitzgerald2022twist,ferreira2022signatures}. For the $\varphi=90^\circ$ case, we find qualitatively the same results (Figs. \ref{fig:perov_tuning}(e)-(g)), with the main distinction from the $0^\circ$ case being the different Hopfield coefficients, i.e., different light-matter compositions of the polariton branches. A plot of the Hopfield coefficients as a function of the angle of incidence can be found in the SI. These results reveal how a combination of magneto-optical measurements and modeling using the Hopfield method and Elliott formula can unravel the interplay of optical selection rules and the balance of decay channels that determine the optical response of exciton polaritons in 2D perovskites.

\section*{Conclusions}

We have revealed the rich magneto-optical response of dark and bright exciton states in 2D perovskites, both in vacuum and when integrated within a Fabry-P\'erot microcavity. In particular, we show that the dark state can enter the strong coupling regime under an applied in-plane magnetic field. Furthermore, we predict anisotropic polariton manifolds in 2D perovskites arising from a magnetic field-induced anisotropy. We also demonstrate that the dispersion and the absorption of different polariton branches can be tuned by adjusting the azimuth angle of the light beam relative to the magnetic field. These gained insights contribute to a deeper microscopic understanding of exciton polaritons in 2D perovskites and could also be important for technologically relevant directional tuning and polarization control of both polariton transport and emission.  

\begin{suppinfo}

Details on the derivation of the main equations and additional discussion of the magnetic-field-dependent polariton landscape (PDF)

\end{suppinfo}
\begin{acknowledgement}
    We acknowledge funding from the DFG via SFB 1083 and the regular project 524612380. We also want to thank Joshua Thompson (University of Cambridge) and Ellie Kraus (Phillips-Universität Marburg) for their valuable discussion as well as Giuseppe Meneghini (Phillips-Universität Marburg) for his help with Figure 1(a).
\end{acknowledgement}

\bibliography{ref}

\end{document}


\maketitle
\section{Methods}
\subsection{Wannier equation}
To obtain exciton energies and wavefunctions, we solve the Wannier equation \cite{brem2018exciton}
\begin{align}
   \sum_{\textbf{k}'}\left(\frac{\hbar^2k'^2}{2 m_r}\delta_{kk'}+V_{|\textbf{k}-\textbf{k}'|}\right)\Psi_n(\textbf{k}')&=E_b^n(\textbf{k})\Psi_n(\textbf{k})\;\text{,}
   \label{eq:wannier_rel}
\end{align}
where $m_r=0.108 m_e$ is the reduced mass of the electron and hole \cite{ziegler2020fast}, $V_q$ is the screened Coulomb potential, $E_n$ are the exciton binding energies, and the corresponding wavefunctions are denoted by $\Psi_n$. The electron and hole masses are taken from Ref.~\citenum{ziegler2020fast} and the Keldysh approximation is used for the screened Coulomb potential \cite{feldstein2020microscopic}. The high-frequency dielectric constants of the organic spacer layer and the inorganic perovskite layer are set to \cite{hong1992dielectric} $3.3$ and $6.1$ , respectively. The thickness of the perovskite slab is set to \cite{hong1992dielectric} $0.636\,\mathrm{nm}$.

\subsection{Exchange interaction and optical selection rules}
Considering only the 1s exciton, the spin combination of the constituent electron and hole leads to four degenerate states. These states interact with each other through the exchange interaction, as well as with an applied magnetic field via the Zeeman effect, leading to the following eigenvalue problem in the exciton basis \cite{thompson2024phonon}
\begin{align}
    \begin{pmatrix}
        I_Z & -g_v(B) & g_c(B) & -I_Z\\
        -g_v(B) & I_r & 0 & g_c(B)\\
        g_c(B) & 0 & I_r & -g_v(B)\\
        -I_Z & g_c(B) & -g_v(B) & I_Z
    \end{pmatrix}\begin{pmatrix}
        D_{\mu,\textbf{q}}^{\uparrow\uparrow}\\
        D_{\mu,\textbf{q}}^{\uparrow\downarrow}\\
        D_{\mu,\textbf{q}}^{\downarrow\uparrow}\\
        D_{\mu,\textbf{q}}^{\downarrow\downarrow}\\
    \end{pmatrix}&=E_{\mu,\textbf{q}}^{(X)}(B)\begin{pmatrix}
        D_{\mu,\textbf{q}}^{\uparrow\uparrow}\\
        D_{\mu,\textbf{q}}^{\uparrow\downarrow}\\
        D_{\mu,\textbf{q}}^{\downarrow\uparrow}\\
        D_{\mu,\textbf{q}}^{\downarrow\downarrow}\\
    \end{pmatrix}\;\text{,}
\end{align}
where $I_{r(Z)}$ quantify both the short- and long-range exchange interaction strength in the in-plane (out-of-plane) direction of the perovskite. Furthermore, $g_{c/v}(B)=g_{c/v}\frac{\mu_B B}{2}$, and $g_{c}=2.9$ and $g_v=-1.1$ denote the g-factors in the excitonic basis \cite{dyksik2021brightening}. At zero magnetic field, the resulting excitonic fine structure states at $|\textbf{q}|=0$ are given by
\begin{align*}
    E_D^{(X)}=\Delta+E_b^{1\mathrm{s}}\;\text{,} \quad E_T^{(X)}=\Delta+E_b^{1\mathrm{s}}+I_r\;\text{,} \quad   E_L^{(X)}=\Delta+E_b^{1\mathrm{s}}+I_r\;\text{,} \quad   E_Z^{(X)}=\Delta+E_b^{1\mathrm{s}}+2I_Z\;\text{,}
\end{align*}
where $\Delta$ is the bandgap energy. The respective eigenvectors are given by
\begin{align*}
    D_D=\frac{1}{\sqrt{2}}\left(1,0,0,1\right)\;\text{,}\quad     D_T=\left(0,1,0,0\right)\;\text{,}\quad    D_L=\left(0,0,1,0\right)\;\text{,}\quad
    D_Z=\frac{1}{\sqrt{2}}\left(1,0,0,-1\right)\;\text{.}
\end{align*}
Here, we fit $I_r$ and $I_Z$ to the experimentally obtained values for the splitting between the dark ($X_D$) and bright states ($X_{T/L}$), as well as between the dark and gray state ($X_Z$) \cite{dyksik2021brightening}.

\begin{figure}[t]
    \centering
    \includegraphics[width=1\textwidth]{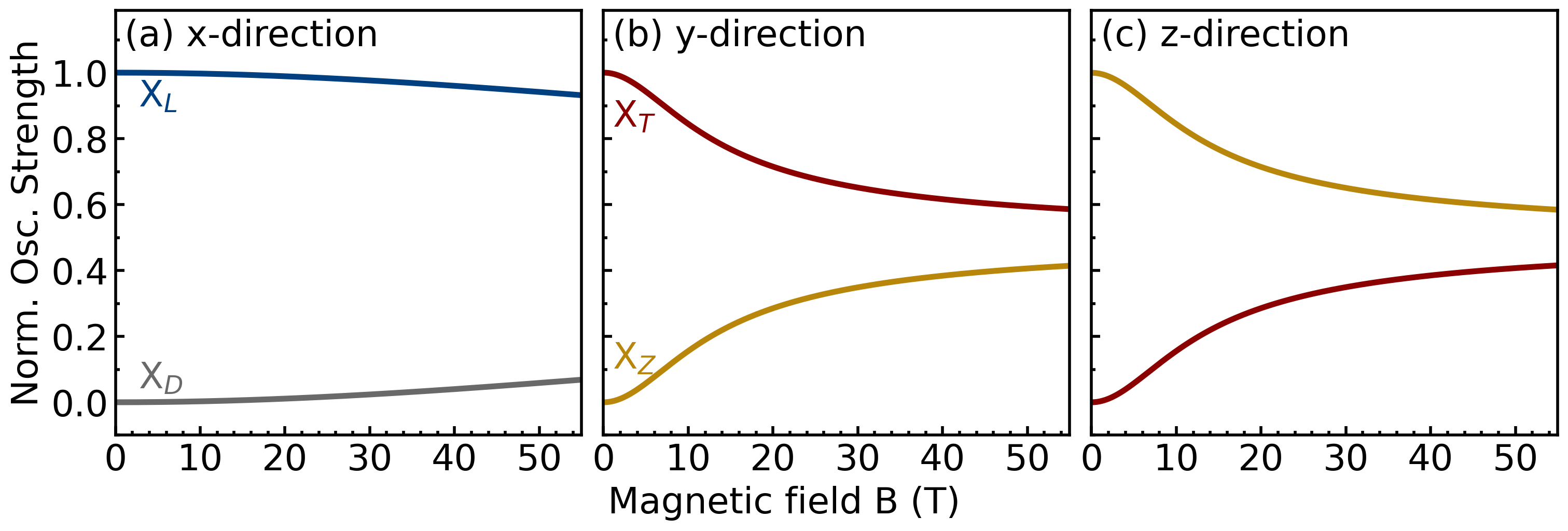}
    \caption{(a) Oscillator strength in the x-direction of the excitonic fine structure states, $X_D$ and $X_L$, polarized in the x-direction (longitudinal to the magnetic field).  (b)-(c) Oscillator strength in the y-direction (transversal and in-plane) and the z-direction (transversal and out-of-plane) of the corresponding states, $X_T$ and $X_Z$, respectively.}
    \label{fig:S_osc}
\end{figure}

The oscillator strength $|M^\mu_\sigma(\textbf{B})|^2$ of each state is a linear combination of the dipole moments of each exciton, $\textbf{d}_{ss'}^{cv}$, weighted by the respective eigenvector, and then projected onto the polarization of the light \cite{thompson2024phonon} $\textbf{e}_\sigma$ 
\begin{align*}
    |M^\mu_\sigma|^2(\textbf{B})\propto |\Psi_{1s}(\textbf{r}=0)|^2|\textbf{e}_\sigma \cdot\sum_{ss'}\textbf{d}_{ss'}^{cv}D_\mu^{ss'}(\textbf{B})\;\text{,}
\end{align*}
where $\Psi_{1s}(\textbf{r})$ is the wavefunction in real space, and $s$ $(s')$ denotes the electron (hole) spin for the respective exciton. Without a magnetic field, the dark exciton state has a zero transition dipole moment, the two degenerate bright states are circularly polarized in the $xy$-plane, and the gray state is polarized along the out-of-plane $z$ direction. The magnetic field mixes these states, leading to the modified selection rules, cf.~Fig. 1 of the main text. The magnetic field dependence of the oscillator strength for different polarization orientations is shown in Fig~\ref{fig:S_osc}. In particular, Fig.~\ref{fig:S_osc}(a) illustrates the transfer of oscillator strength from the bright state $X_L$ to the dark exciton $X_D$ with increasing magnetic field. Figures \ref{fig:S_osc}(b) and (c) show the mixing of the oscillator strength for the transverse state $X_T$ and the out-of-plane gray exciton $X_Z$, due to the rotation of the respective transition dipole moments around the magnetic field axis (Fig.~1 in the main text). Note that for all magnetic fields, the total oscillator strength in each polarization direction is equal to unity.

\subsection{Scattering matrix method}
The scattering matrix (S-matrix) method is a numerical algorithm used to exactly solve Maxwell's equations for stacks of dielectric slabs with in-plane translational invariance or periodicity \cite{rumpf2011improved}. It provides access to the optical response of the system for any incident light angle. However, it is difficult to disentangle photonic and material contributions of a mixed light-matter state, as described by Hopfield coefficients. In particular, this is necessary to obtain access to the different decay channels associated with the constituent excitons and cavity modes. To this end, we use a Hopfield model (see Eq.~3 in the main text) using parameters extracted from S-matrix simulations.

\begin{figure}[b!]
    \centering
    \includegraphics[width=0.66\textwidth]{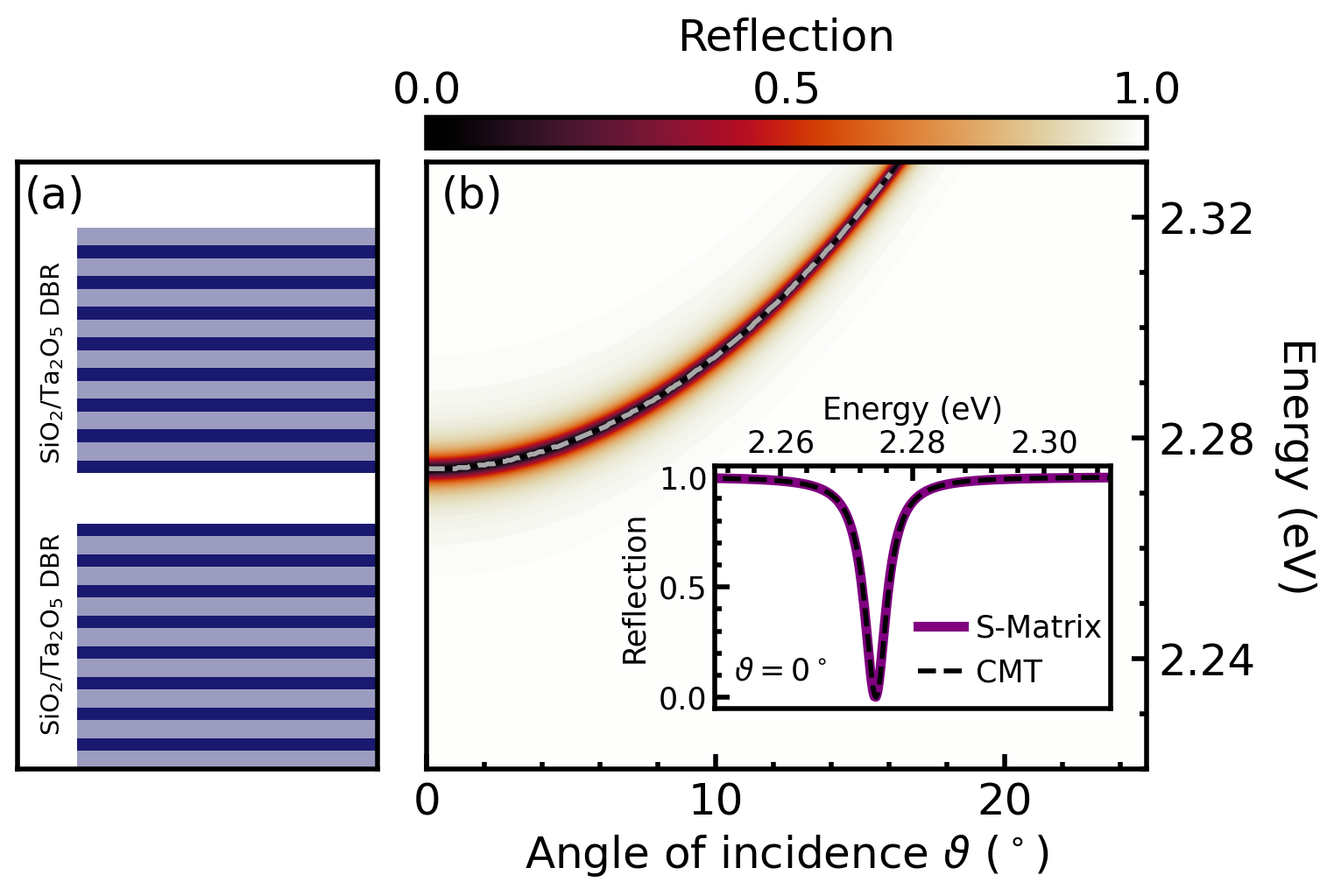}
    \caption{(a) Schematic of the cavity used for the S-matrix simulations, constructed from two $\lambda/4$-DBR mirrors, each consisting of alternating stacked SiO$_2$ (light blue) and TaO$_2$ (dark blue) layers. (b) S-matrix simulation of the reflection of the bare Fabry-Perot microcavity as a function of the photon energy and angle of incidence $\vartheta$. The gray dashed line illustrates the extracted cavity photon mode energy, perfectly following the peak reflection. The inset shows a line cut at $\vartheta=0^\circ$ (purple solid line) and the fit using coupled mode theory (black dashed line) with the extracted photon linewidth $\hbar\kappa$.}
    \label{fig:S_bare_sim}
\end{figure}

In this work, we consider a microcavity consisting of two $\lambda/4$-DBR mirrors, each consisting of eight stacks of alternating SiO$_2$ and TaO$_2$ ($n_{\mathrm{SiO}_2}=1.46$ and $n_{\mathrm{Ta}_2\mathrm{O}_5}=2.0770$ \cite{rodriguez2016self}) dielectric slabs, for a total number of 16 layers per DBR. A sketch of the simulated cavity can be seen in Fig.~\ref{fig:S_bare_sim}(a). The thickness of the layers was chosen such that the center frequency of the DBR aligns with the energy of the dark state. To simulate and extract the cavity mode as a function of angle of incidence, we also have to include the dielectric background of the perovskite layer in the center of the cavity, i.e., without any excitonic effects. This is because a thin dielectric slab will affect the resonance condition of the microcavity. Using a dielectric background for the perovskite layer of 1.81 \cite{fieramosca2018tunable}, we obtain the reflection of the cavity, as shown in Fig.~\ref{fig:S_bare_sim}(b). By fitting a Lorentzian function to the reflection (Eq.~\ref{eq:S_CM_R} in the limit of a bare cavity), we are able to obtain the cavity linewidth $\hbar\kappa$, see the inset of Fig.~\ref{fig:S_bare_sim}(b). Including the dielectric tensor given in Eqs.~(1) and (2) of the main text, we can calculate the optical response of the (PEA)$_2$PbI$_4$ perovskite layer integrated into the microcavity, and then extract the polariton energies from the dips in reflection. These energies are then fit to the Hopfield model in Eq.~(3) of the main text to obtain access to the cavity photon-exciton coupling strengths and Hopfield coefficients.

\subsection{Coupled mode theory}
\label{sec:CMT}

Assuming high Q-factor cavity modes and small material-based losses, coupled mode theory (CMT) provides an intuitive and simple description of the coupling between different modes of the system (excitons, cavity photons) and their coupling to external ports \cite{fan2003temporal}. In particular, it provides insight into calculated absorption spectra via the critical coupling condition \cite{ferreira2022signatures,konig2023interlayer,ferreira2024revealing}. Here, we detail the derivation of the polaritonic Elliott formula using classical two-port, two-resonator CMT equations. Starting from the Hamiltonian in Eq. (3) of the main text, we add an imaginary part to the exciton and cavity-photon energies on the diagonal to describe the respective loss parameters. We then obtain the following coupled dynamics of excitons and cavity photons:
\begin{align*}
   \partial_t\begin{pmatrix}
        C(t)\\X_1(t)\\X_2(t)
    \end{pmatrix}&=-\frac{i}{\hbar}H\cdot\begin{pmatrix}
        C(t)\\X_1(t)\\X_2(t)
    \end{pmatrix}=\begin{pmatrix}
        -i\omega^{(C)}+\kappa & -i\frac{g_1}{\hbar} & -i\frac{g_2}{\hbar}\\
          -i\frac{g_1}{\hbar} & -i\omega_1^{(X)}+\Gamma & 0\\
           -i\frac{g_2}{\hbar} & 0 & -i\omega_2^{(X)}+\Gamma
    \end{pmatrix}\cdot\begin{pmatrix}
        C(t)\\X_1(t)\\X_2(t)
    \end{pmatrix}
  \end{align*}
Here, $X_\mu(t)$ is the mode amplitude of the $\mu$th exciton oscillating in time with the frequency $\omega_\mu^{(X)}$ and decay rate $\Gamma$. Furthermore, $C(t)$ is the amplitude of the cavity photon at the frequency $\omega^{(C)}$ with the decay rate $\kappa$. Using the Hopfield transformation \cite{hopfield1958theory,fitzgerald2022twist} $C(t)=\sum_n P_n(t) U_0^n$ and $X_\mu(t)=\sum_n P_n(t) U_\mu^{n}$, we can decouple the dynamics into a set of independent differential equations. Here, $P_n(t)$ is the polariton mode amplitude of the $n$th branch with $U_0^{n}$ and $U_\mu^{n}$ as the respective photonic and excitonic Hopfield coefficients.

\begin{figure}[b!]
    \centering
    \includegraphics[width=0.5\textwidth]{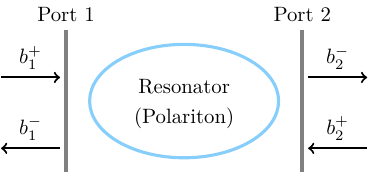}
    \caption{Schematic of the ports 1 and 2 coupling the respective incoming waves ($b_1^+$, $b_2^+$) and outgoing waves ($b_1^-$, $b_2^-$) to the resonator, i.e., the polariton.}
    \label{fig:S_ports}
\end{figure}

Next, we add the coupling to external photons, i.e., the two ports corresponding to the continuum of photon states in the half-space either side of the cavity \cite{fan2003temporal,fitzgerald2022twist}. In a Fabry-Perot microcavity, only the photonic component of each polariton branch can couple to the ports, i.e., excitons are not directly excited by external photons. The incoming and outgoing waves are given by
\begin{align}
    \left(b^+\right)^T(t)=\left(b_1^+(t),b_2^+(t)\right), \quad  \left(b^-\right)^T(t)=\left(b_1^-(t),b_2^-(t)\right)\;\text{,}
\end{align}
where $b_{i}^{-(+)}$ is the incoming (outgoing) wave at the $i$th port, as shown in Fig.~\ref{fig:S_ports}. The incoming waves couple to the photonic part of the polaritons through 
$\xi^T_n=\left(\xi_{n,1},\xi_{n,2}\right)$ for each port respectively, while the outgoing waves couple via $d^T_n=\left(d_{n,1},d_{n,2}\right)$. The direct process of the incoming waves coupling to the outgoing waves is described by a $2\times2$ matrix $\mathcal{C}$. As direct transmission is not possible in an idealized Fabry-Perot cavity, we can only couple $b_{i}^+$ to $b_{i}^-$ of the same port via reflection. Therefore, the matrix $\mathcal{C}$ is the negative of the identity matrix. Note that one can also add the coupling to the ports before performing the Hopfield transformation and obtain the same equations.

Exploiting time-reversal symmetry \cite{fan2003temporal,zhao2019connection}, which is valid for low absorptive losses, we obtain 
\begin{align*}
\xi_n=d_n=
\begin{pmatrix}
\sqrt{2\gamma_{n,1}^{(P)}}\\\sqrt{2\gamma_{n,2}^{(P)}}
\end{pmatrix}\;\text{,}
\end{align*} 
where $\gamma_{n,i}^{(P)}$ is the photonic decay rate of the $n$th polariton branch into the $i$th port. The sum of the polaritonic radiative decay rate into each port is equal to the total photonic-based polariton decay rate from the main text, i.e., $\gamma_{n,1}^{(P)}+\gamma_{n,2}^{(P)}=\gamma_n^{(P)}$.
This results in the polariton dynamics
\begin{subequations}
\begin{align}
    \partial_t P_n(t)&=\left(-i\omega_n-\gamma_n^{(P)}-\Gamma_n^{(P)}\right)P_n(t)+
        \sqrt{2\gamma_{n,1}^{(P)}}b^+_1(t)+\sqrt{2\gamma_{n,2}^{(P)}}b^+_2(t)\;\text{,}\label{eq:S_CM1}\\[10pt]  
    \begin{pmatrix}
        b^-_1(t)\\b^-_2(t)
    \end{pmatrix}&=\begin{pmatrix}
        -b^+_1(t)\\
        -b^+_2(t)
    \end{pmatrix}+\sum_n\begin{pmatrix}\sqrt{2\gamma_{n,1}^{(P)}}\\\sqrt{2\gamma_{n,1}^{(P)}}\end{pmatrix}P_n(t)\;\text{,}\label{eq:S_CM2}
\end{align} 
\end{subequations}
where $\omega_n$ and $\Gamma_n^{(P)}$ are the frequency and material-based decay rates of the $n$th polariton branch, and $b_i^{+(-)}$ are the incoming (outgoing) fields of the $i$th port. The first equation describes how polaritons are excited by the incoming waves, while the second equation expresses the emission of the outgoing waves.

As we consider only excitations from one port, we set $b_2^+(t)=0$. For well-spaced polaritons, we can ignore any overlap and therefore solve the equations independently for one polariton branch, and then simply add the final contributions together to obtain the total linear optical spectra. The Fourier transformation of Eqs.~\ref{eq:S_CM1} and \ref{eq:S_CM2} gives the reflection and transmission coefficients for each polariton branch:
\begin{align*}
    r_n(\omega)&=\frac{b_1^-(\omega)}{b_1^+(\omega)}=\frac{-i(\omega-\omega_n)+\left(\gamma_{n,1}^{(P)}-\gamma_{n,2}^{(P)}-\Gamma_n^{(P)}\right)}{i(\omega_n-\omega)+\gamma_n^{(P)}+\Gamma_n^{(P)}}\;\text{,}\\[10pt]
    it_n(\omega)&=\frac{b_2^-(\omega)}{b_1^+(\omega)}=\frac{2\sqrt{\gamma_{n,1}^{(P)}\gamma_{n,2}^{(P)}}}{i(\omega_n-\omega)+\gamma_n^{(P)}+\Gamma_n^{(P)}}\;\text{,}
\end{align*}
In the limit of a symmetric cavity, the reflection, transmission, and absorption for each polariton branch are then given by
\begin{subequations}
    \begin{align}
        R_n(\omega)&=\frac{(\omega_n-\omega)^2+\left(\Gamma_n^{(P)}\right)^2}{(\omega_n-\omega)^2+\left(\gamma_n^{(P)}+\Gamma_n^{(P)}\right)^2}\;\text{,}\label{eq:S_CM_R}\\[10pt]
        T_n(\omega)&=|it(\omega)|^2=\frac{\left(\gamma_{n}^{(P)}\right)^2}{(\omega_n-\omega)^2+\left(\gamma_n^{(P)}+\Gamma_n^{(P)}\right)^2}\;\text{,}\label{eq:S_CM_T}\\[10pt]
        A_n(\omega)&=1-R(\omega)-T(\omega)=\frac{2\gamma_{n}^{(P)}\Gamma_n^{(P)}}{(\omega_n-\omega)^2+\left(\gamma_n^{(P)}+\Gamma_n^{(P)}\right)^2}\;\text{,}\label{eq:S_CM_A}    
    \end{align}
\end{subequations}
resulting in the polaritonic Elliott formula (Eq. 4 of the main text). These equations also apply to the case of a bare Fabry-Perot cavity by setting $\Gamma_n^{(P)}=0$ and replacing $\gamma_n^{(P)}$ with the bare cavity decay rate $\kappa$. This allows us to fit and extract the cavity photon linewidth from the S-matrix simulation, as shown by the black dashed curve in the inset of Fig.\ref{fig:S_bare_sim}. Furthermore, in the excitonic limit, $ \Gamma^{(P)} = \Gamma$ and ${\gamma}^{(P)}=\gamma$, with $\gamma$ being the radiative decay rate of the exciton, equation \ref{eq:S_CM_A} gives the well-known excitonic Elliott formula \cite{kira2006many}. In this case, excitons can couple directly to the ports, rather than indirectly via cavity photons. The Elliott formula reveals that the maximum absorption of 0.5 at resonance is reached when the critical coupling condition $\Gamma_n^{(P)}=\gamma_n^{(P)}$ is met \cite{fitzgerald2022twist,ferreira2022signatures,konig2023interlayer,ferreira2024revealing}.

\begin{figure}[t!]
    \centering
    \includegraphics[width=0.6\textwidth]{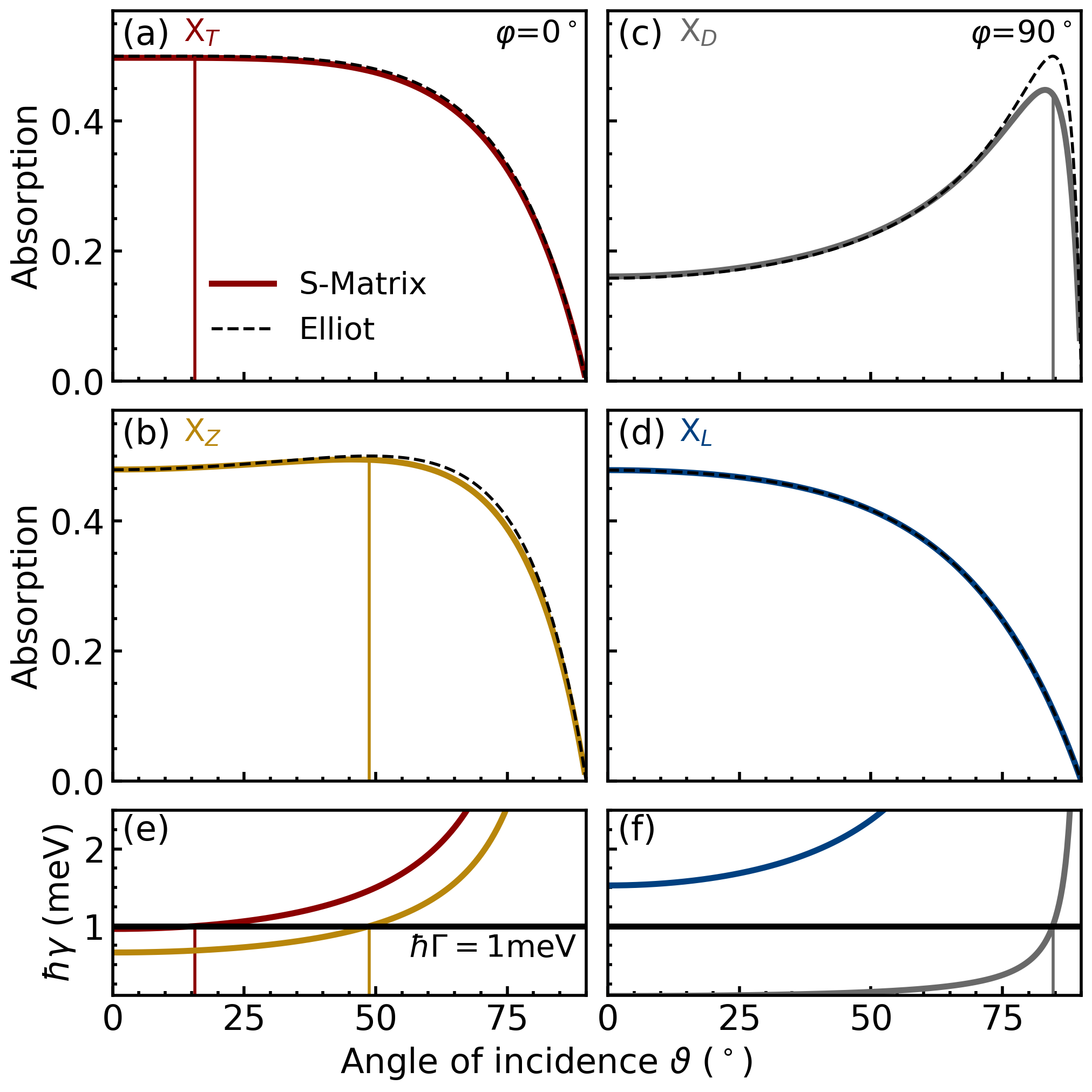}
    \caption{(a)-(d) Absorption of a (PEA)$_2$PbI$_4$ perovskite layer as a function of the angle of incidence, $\vartheta$, resonant with the four exciton fine structure states $X_T$, $X_Z$, $X_D$ and $X_L$ calculated using the S-matrix method (colored solid lines) and the Elliott formula (black dashed lines). (e)-(f) Radiative decay of the respective states (colored lines) and material-based loss (black horizontal line) as a function of $\vartheta$ for the states shown in (a)-(d), respectively. The colored vertical lines illustrate, where the excitonic decay rate equals the respective photonic decay, i.e., where the critical coupling condition is met resulting in a maximum absorption.}
    \label{fig:S_SM_Elliott_comp}
\end{figure}

Comparing the absorption of the single perovskite layer between the S-matrix method and the Elliott formula in Fig.~\ref{fig:S_SM_Elliott_comp}, we find excellent agreement for all states. The minor deviation reflects a small spectral overlap between the $X_T$ and $X_Z$ excitons, which is not taken into account in the Elliott formula. Additionally, we find that the critical coupling condition fully describes the absorption behavior, as illustrated by the vertical colored lines in Fig. \ref{fig:S_SM_Elliott_comp}.

\begin{figure}[t]
    \centering
    \includegraphics[width=0.5\textwidth]{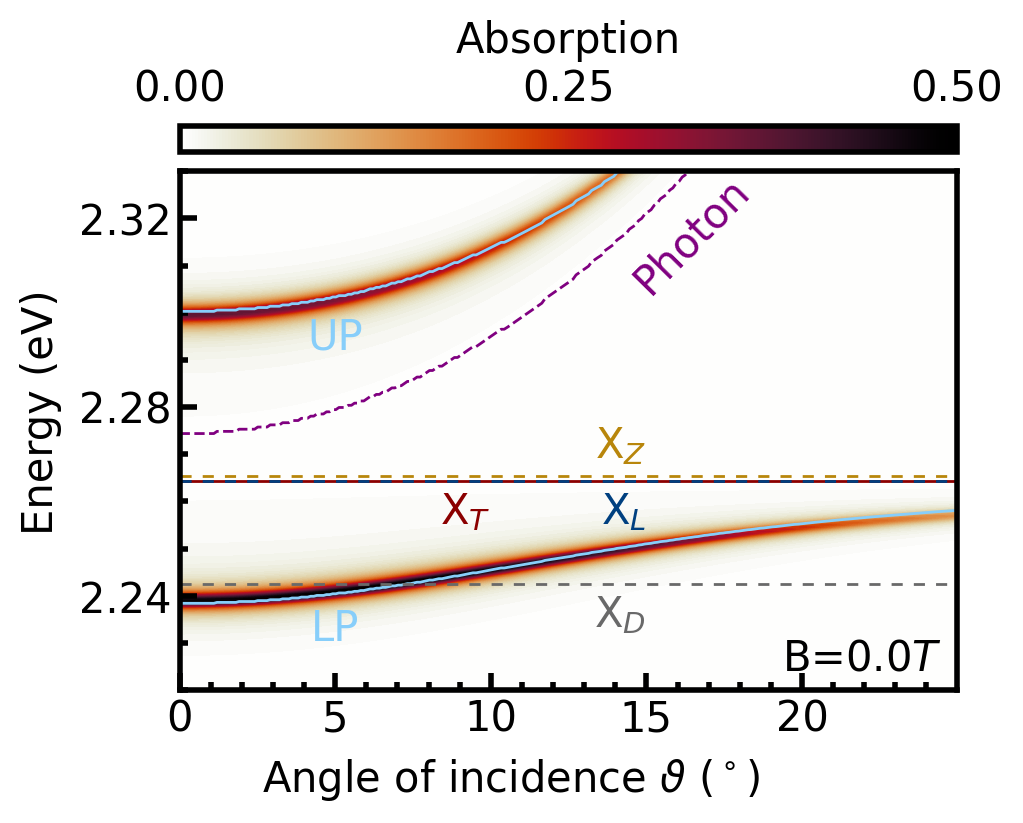}
    \caption{Absorption of a 2D (PEA)$_2$PbI$_4$ perovskite layer integrated within a Fabry-Perot microcavity as a function of the photon energy and angle of incidence at $B=0\,\mathrm{T}$. The dashed horizontal lines denote the energy of the different excitonic states. As only the degenerate bright states $X_T$ and $X_L$ can couple to the light (due to the selection rules shown in Fig. \ref{fig:S_osc}), only two polariton branches, LP and UP, are visible with a fully dark middle polariton branch lying on top of the two degenerate states. As there is no breaking of in-plane symmetry, the absorption spectrum is independent of the azimuth angle.}
    \label{fig:S_pol_woB}
\end{figure}

\section{Polariton landscape}
Without a magnetic field, only the two circularly polarized bright states $X_T$ and $X_L$ can couple to the TE-polarized light, as shown in Figs.~\ref{fig:S_osc}(a)-(b). Therefore, only two azimuth-angle-independent polariton branches are visible in absorption, cf. Fig.~\ref{fig:S_pol_woB}. In other words, the polariton dispersion is isotropic in the absence of a magnetic field. These two branches are equivalent to the polaritons visible in Fig.~3(a) of the main text, showing a single large Rabi splitting centered on $X_T$. This is because even at $B=50\,\mathrm{T}$, the two transversally polarized states $X_T$ and $X_Z$ are still almost degenerate in energy and have together the same in-plane oscillator strength as $X_T$ at $B=0\,\mathrm{T}$ due to the conservation of oscillator strength, cf. Fig. \ref{fig:S_osc}(b).

\begin{figure}[t!]
    \centering
    \includegraphics[width=1\textwidth]{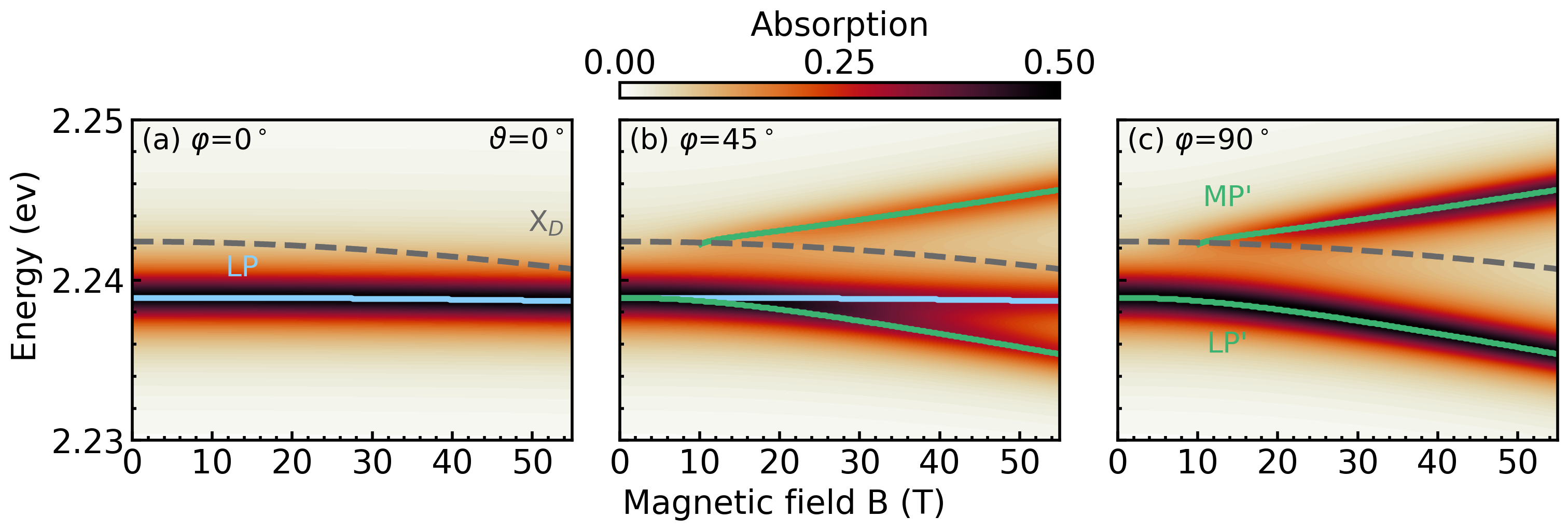}
    \caption{(a)-(c) Absorption of a 2D (PEA)$_2$PbI$_4$ perovskite layer integrated within a Fabry-Perot microcavity as a function of the photon energy and magnetic field strength $B$, for $\vartheta=0^\circ$, and for three different azimuth angles $\varphi=0^\circ$, $45^\circ$, and $90^\circ$. The dashed line denotes the energy of the dark state $X_D$. As the latter can not couple to light at $\varphi=0^\circ$, according to the selection rules, only a single polariton branch (LP) stemming from the energetically higher bright states can be observed in this energy range. In contrast, for $\varphi=90^\circ$, the dark state can couple to the cavity mode, and therefore two branches appear (LP' and MP').}
    \label{fig:S_pol_Bsweep}
\end{figure}
The dependence of the polariton landscape on the magnetic field strength is shown in Fig.~\ref{fig:S_pol_Bsweep}. Similar to Fig.~3 in the main text, it reveals a single branch (LP) around the energy of the dark state at $\varphi=0^\circ$, two at $\varphi=90^\circ$ (LP' and MP') and a superposition of the two edge cases at $\varphi=45^\circ$ with approximately halved absorption. The MP' branch can only be resolved starting at about 10\,T because, below this field strength, the dark state lacks sufficient oscillator strength to enter the strong coupling regime, i.e., $g<(\hbar\gamma+\hbar\Gamma)/2$. The energies of LP' and MP' polaritons are used to calculate the Rabi splitting around the dark state as a function of $B$, shown in Fig. 4(a) of the main text.

\section{Hopfield coefficients}
\begin{figure}[h!]
        \centering
        \includegraphics[width=0.6\textwidth]{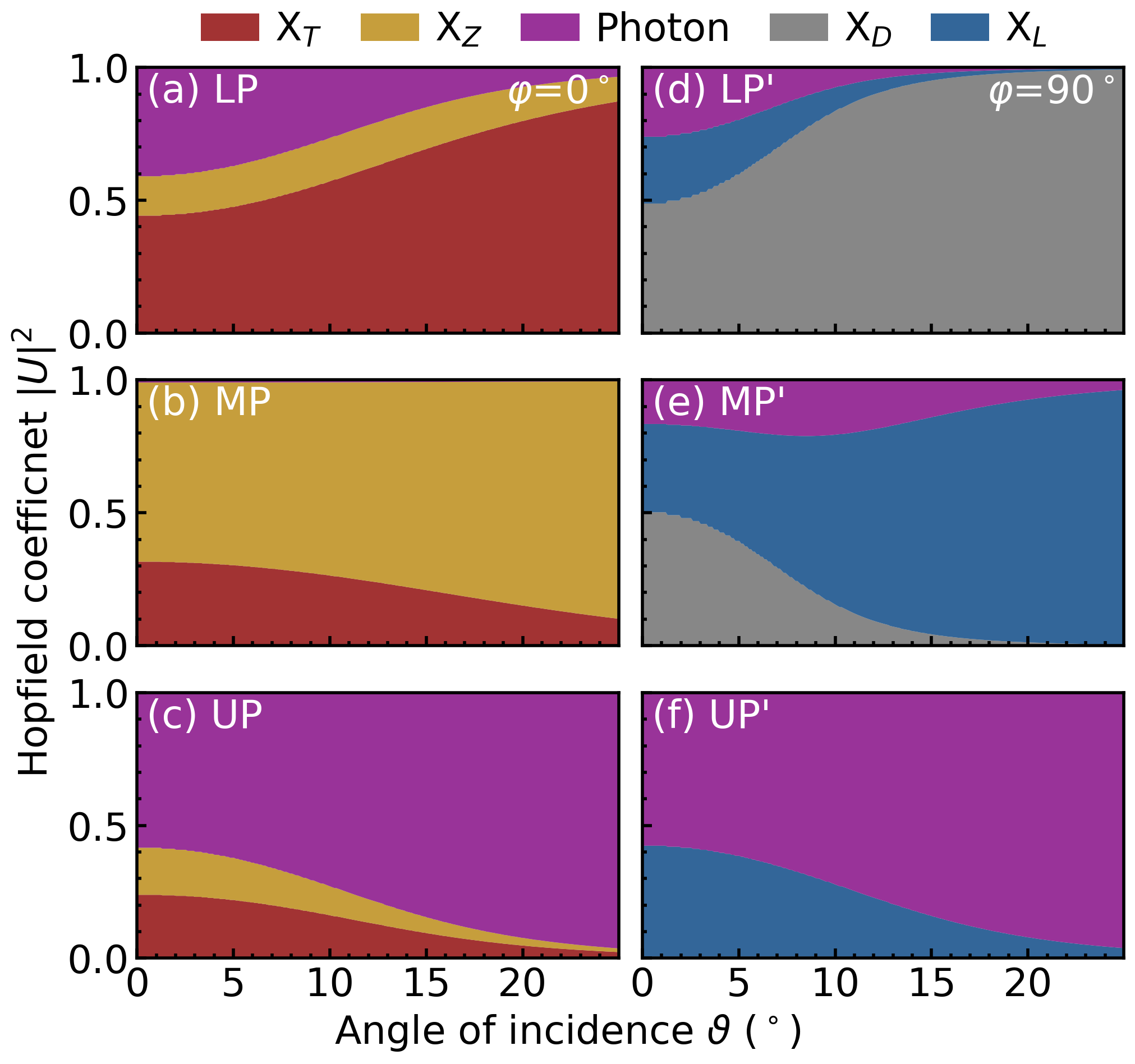}
        \caption{Hopfield coefficients as a function of angle of incidence, $\vartheta$, for the polariton branches shown in Figs. 3(a) and (c) of the main text. They describe the excitonic and photonic composition of these branches.}
        \label{fig:S_Hop}
    \end{figure}

The Hopfield coefficients for the polariton branches shown in Fig.~\ref{fig:S_Hop} of the main text are obtained by fitting the Hopfield/CMT model to the reflection spectrum. Both lower branches, LP and LP', have the highest contribution from their respective excitons in the small-incident-angle limit, while in the high-angle limit, they become almost fully photonic. The two upper branches, UP and UP' show the opposite behavior: they possess the highest photonic contribution at small angles and are almost fully excitonic at high angles \cite{fitzgerald2022twist}. For $\varphi=0^\circ$, the middle polariton branch, MP, has a negligible photonic component and therefore appears flat and almost dark, cf. Fig. 3(a) of the main text. In contrast, the MP' branch for $\varphi=90^\circ$ exhibits a photonic component and thus has a sizable absorption (Fig.~3(c) of the main text). These Hopfield coefficients at normal incidence, $\vartheta=0^\circ$, are used to obtain the excitonic and radiative linewidths $\hbar\Gamma_n^{(P)}$ and $\hbar\gamma^{(P)}$, which help further understand the absorption shown in Fig.~4 of the main text.

\bibliography{ref}